\documentclass[twocolumn]{aastex62}

\usepackage{graphicx}
\usepackage{amsmath}	
\usepackage{amssymb}
\usepackage{booktabs}
\usepackage{enumerate}
\usepackage{dirtytalk}
\usepackage{float}
\usepackage{enumitem}
\usepackage{booktabs}
\usepackage{lipsum} 
\usepackage{float}
\usepackage{xcolor}
\usepackage{multirow}
\usepackage{makecell}
\usepackage{blindtext}
\usepackage{chngcntr}

\def\Mj{\,{\rm$\text{M}_\text{J}$}}

\graphicspath{{./}{figures/}}

\received{August 11, 2019}
\accepted{November 21, 2019}

\submitjournal{ApJL}

\shorttitle{Mass metallicity trends in exoplanets}
\shortauthors{Welbanks et al.}

\begin{document}

\title{Mass-Metallicity Trends in Transiting Exoplanets from Atmospheric Abundances of H$_2$O, Na, and K}

\author{Luis Welbanks}
\email{luis.welbanks@ast.cam.ac.uk}
\affil{Institute of Astronomy, University of Cambridge, Madingley Road Cambridge CB3 0HA, UK}

\author{Nikku Madhusudhan}
\email{nmadhu@ast.cam.ac.uk}
\affil{Institute of Astronomy, University of Cambridge, Madingley Road Cambridge CB3 0HA, UK}

\author{Nicole F. Allard}
\affil{GEPI, Observatoire de Paris PSL Research University, UMR 8111, CNRS, Sorbonne Paris Cit\'e,61, Avenue de l'Observatoire, F-75014 Paris, France }
\affil{Institut d'Astrophysique de Paris, UMR7095, CNRS, Universit\'e Paris VI, 98bis Boulevard Arago, PARIS, France}

\author{Ivan Hubeny}
\affil{Department of Astronomy, University of Arizona, Tucson AZ 85721, USA}

\author{Fernand Spiegelman}
\affil{Laboratoire de Chimie et de Physique Quantiques, Universit\'e de Toulouse (UPS) and CNRS, 118 route de Narbonne, F-31400 Toulouse, France}

\author{Thierry Leininger}
\affil{Laboratoire de Chimie et de Physique Quantiques, Universit\'e de Toulouse (UPS) and CNRS, 118 route de Narbonne, F-31400
Toulouse, France}

\begin{abstract}
Atmospheric compositions can provide powerful diagnostics of formation and migration histories of planetary systems. We investigate constraints on atmospheric abundances of H$_2$O, Na, and K, in a sample of transiting exoplanets using latest transmission spectra and new H$_2$ broadened opacities of Na and K.  Our sample of 19 exoplanets spans from cool mini-Neptunes to hot Jupiters, with equilibrium temperatures between $\sim$300 and 2700~K. Using homogeneous Bayesian retrievals we report atmospheric abundances of Na, K, and H$_2$O, and their detection significances, confirming 6 planets with strong Na detections, 6 with K, and 14 with H$_2$O.  We find a mass-metallicity trend of increasing H$_2$O abundances with decreasing mass, spanning generally substellar values for gas giants and stellar/superstellar for Neptunes and mini-Neptunes. However, the overall trend in H$_2$O abundances, from mini-Neptunes to hot Jupiters, is significantly lower than the mass-metallicity relation for carbon in the solar system giant planets and similar predictions for exoplanets. On the other hand, the Na and K abundances for the gas giants are stellar or superstellar, consistent with each other, and generally consistent with the solar system metallicity trend. The H$_2$O abundances in hot gas giants are likely due to low oxygen abundances relative to other elements rather than low overall metallicities, and provide new constraints on their formation mechanisms. The differing trends in the abundances of species argue against the use of chemical equilibrium models with metallicity as one free parameter in atmospheric retrievals, as different elements can be differently enhanced.
\end{abstract}

\keywords{methods: data analysis --- planets and satellites: composition --- planets and satellites: atmospheres}

\section{Introduction}
\label{sec:intro}
Exoplanet science has entered an era of comparative studies of planet populations. Several studies have used empirical metrics for comparative characterization of giant exoplanetary atmospheres based on their transmission spectra \citep[e.g.,][]{Sing2016, Stevenson2016a, Heng2016, Fu2017}. Comparative studies are also being carried out using full atmospheric retrievals, primarily constraining H$_2$O abundances and/or cloud properties from transmission spectra \citep[e.g.,][]{Madhusudhan2014a, Barstow2017, Pinhas2019}. Besides H$_2$O, Na and K are the most observed chemical species in giant exoplanetary atmospheres using space- and ground-based telescopes \citep[e.g.,][]{Charbonneau2002, Redfield2008, Wyttenbach2015, Sing2016, Nikolov2018}. As observations improve in precision, recent studies have begun to retrieve Na and K abundances from transmission spectra \citep[e.g][]{Nikolov2018, Pinhas2019, Fisher2019}.

Previous ensemble studies have focused on H$_2$O and found low abundances compared to solar system expectations \citep[e.g.,][]{Madhusudhan2014a, Barstow2017, Pinhas2019}. However, it has been unclear if the low H$_2$O abundances are due to low overall metallicities, and hence low oxygen abundances, or due to high C/O ratios \citep{Madhusudhan2014a} or some other mechanism altogether. Therefore, abundance estimates of other elements such as Na and K provide an important means to break such degeneracies, and provide potential constraints on planetary formation mechanisms \citep[e.g.,][]{Oberg2011,Madhusudhan2014b,Thorngren2016,Mordasini2016}. In the present work, we conduct a homogeneous survey of Na, K, and H$_2$O abundances for a broad sample of transiting exoplanets, and investigate their compositional diversity.

\section{Observations}
\label{sec:observations}

 We consider transmission spectra of 19 exoplanets with masses ranging from 0.03 to 2.10\Mj$\,$ and equilibrium temperatures from 290 to 2700K, as shown in Table~\ref{table:summary}. The spectral range covered in the observations generally spans $0.3\text{-}5.0\mu$m obtained with multiple instruments, including \textit{Hubble Space Telescope} (HST) \textit{Space Telescope Imaging Spectrograph} (STIS) ($\sim$0.3-1.0$\mu$m), HST WFC3 G141 ($\sim$1.1-1.7$\mu$m), and Spitzer photometry (3.6 and 4.5$\mu$m), for most planets and ground-based optical spectra ($\sim$0.4-1.0$\mu$m) for some planets. We select the sample of 10 hot Jupiters with HST and Spitzer observations from \cite{Sing2016}. We expand the sample by including five other planets that have ground-based transmission spectra in the optical: GJ3470b \citep{Chen2017, Benneke2019a}, HAT-P-26b \citep{Stevenson2016b, Wakeford2017}, WASP-127b \citep{Chen2018}, WASP-33b \citep{vonEssen2019}, and WASP-96b \citep{Nikolov2018}. We include four more planets with strong H$_2$O detections: WASP-43b \citep{Kreidberg2014, Stevenson2017},  WASP-107b \citep{Spake2018}, K2-18b \citep{Benneke2019b}, and HAT-P-11b \citep{Chachan2019}.
 
 For the sample of \cite{Sing2016} we use the data selection from \cite{Pinhas2019}, with the exception of WASP-39b for which we use the combined transmission spectrum of \cite{Kirk2019}, and WASP-19b for which we use a ground-based transmission spectrum from Very Large Telescope  \citep[VLT;][]{Sedaghati2017}. The VLT spectrum is of higher resolution than the STIS spectrum and showed evidence for spectral features; we note that the same features were not seen by \cite{Espinoza2019} in a ground-based spectrum obtained with GMT at a different epoch, albeit with lower resolution. For each data set, we follow the same data treatment for retrieval as in the corresponding work. 
 
 The spectral range of the data allow simultaneous constraints on Na, K, and H$_2$O, along with other atmospheric properties. The optical range probes the prominent spectral features of Na ($\sim$589 nm) and K ($\sim$770nm), and contributions from scattering phenomena such as Rayleigh scattering and clouds/hazes \citep[e.g.,][]{Sing2016}, as well as absorption from other chemical species such as TiO \citep[e.g.,][]{Sedaghati2017}. On the other hand, the HST WFC3 and Spitzer bands probe molecular opacity from volatile species such as H$_2$O, CO, and HCN \citep{Madhusudhan2012}. 

\section{H$_2$ broadened Alkali Cross-sections}
\label{sec:broadening}
Given our goal of constraining abundances of Na and K based on optical transmission spectra it is important to ensure accurate absorption cross-sections of these species in the models. We use the latest atomic data on Na and K absorption including broadening due to H$_2$ which is the dominant species in gas giant atmospheres. The Na line data is obtained from \cite{Allard2019} and the K line data is obtained from \cite{Allard2016}. The line profiles are calculated in a unified line shape semiclassical theory \citep{Allard1999} that accounts both for the centers of the spectral lines and their extreme wings, along with accurate ab initio potentials and transition moments.

\begin{figure}
    \centering
    \includegraphics[width=0.5\textwidth]{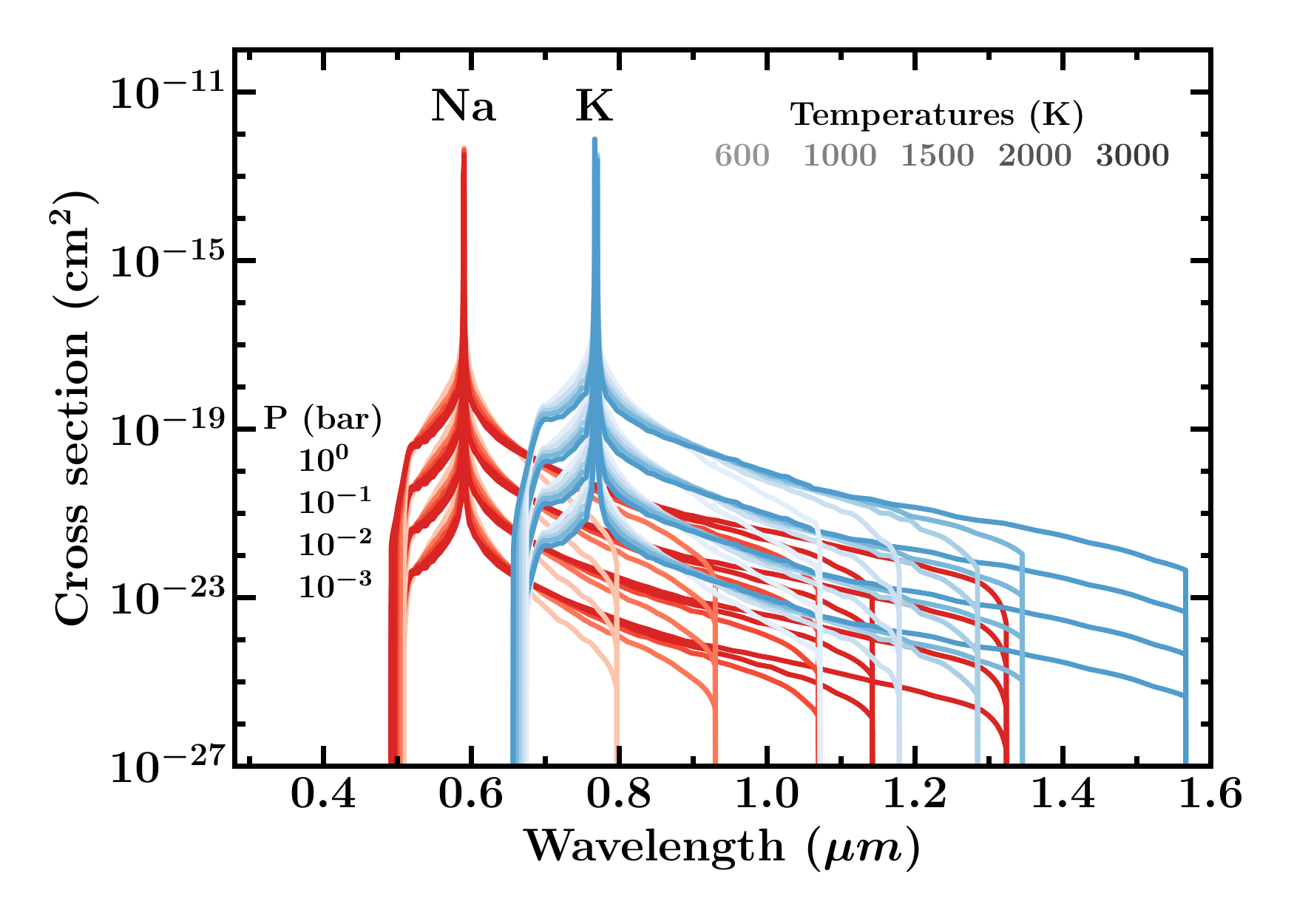}
    \caption{Absorption cross-sections of Na and K broadened by H$_2$ at different pressures and temperatures. Line profiles for each pressure appear as a group. In each group, the darker colors (broader wings) denote hotter temperatures.}
    \label{fig:cross_sec}
\end{figure}

We compute the cross section for H$_2$ broadened Na for both the D1 and D2 doublets at 5897.56\AA$\,$ and 5891.58\AA$\,$ respectively. We calculate the contributions from the core of the lines and their broadened wings for 500, 600, 725, 1000, 1500, 2000, 2500, and 3000K for pressures equally spaced in log space from 10$^{-5}$ to 10$^{2}$ bar. We repeat this procedure for H$_2$ broadened K for the D1 and D2 doublet peaks at 7701.10\AA$\,$ and 7667.02\AA$\,$, at 600, 1000, 1500, 2000, and 3000K. Figure~\ref{fig:cross_sec} shows the H$_2$ broadened cross-sections of Na and K for a range of pressures and temperatures. The extended wings of our alkali cross-sections, particularly of K, stop below $\sim 1.4 \mu$m for the typical pressures and temperatures probed by transmission spectra. As such, these cross-sections do not provide an extended continuum to the spectrum in the HST WFC3 band, which results in retrieved H$_2$O abundances that are conservatively higher. Future calculations including other transition lines in the near infrared and their H$_2$ broadening could extend Na/K opacity into the WFC3 range.
 
\newpage
\section{Atmospheric Retrieval}
\label{sec:retrievals}

We follow the retrieval approach in \cite{Welbanks2019} based on the AURA retrieval code \citep{Pinhas2018}. The code allows for the retrieval of chemical abundances, a pressure-temperature profile, and cloud/haze properties using spectra from multiple instruments. The model computes line by line radiative transfer in a transmission geometry and assumes hydrostatic equilibrium and uniform chemical volume mixing ratios in the atmosphere. A full description of the retrieval setup can be found in \cite{Pinhas2019} and \cite{Welbanks2019}. 

Besides the new H$_2$ broadened alkali species discussed in section~\ref{sec:broadening}, we include opacities due to other chemical species possible in hot giant planet atmospheres \citep{Madhusudhan2012}. Our retrievals generally consider absorption due to H$_2$O \citep{Rothman2010}, Na \citep{Allard2019}, K \citep{Allard2016}, CH$_4$ \citep{Yurchenko2014}, NH$_3$ \citep{Yurchenko2011}, HCN \citep{Barber2014}, CO \citep{Rothman2010}, and H$_2$-H$_2$ and H$_2$-He collision induced absorption \citep[CIA;][]{Richard2012}. Additionally, following previous studies \citep{Chen2018, vonEssen2019, Sedaghati2017, MacDonald2019}, we include absorption due to Li \citep{Kramida2018} for WASP-127b, AlO \citep{Patrascu2015} for WASP-33b, TiO \citep{Schwenke1998} for WASP-19b, and CrH \citep{Bauschlicher2001}, TiO, and AlO for HAT-P-26b. We exclude absorption due to Na and K for K2-18b, GJ-3470b, and WASP-107b as it is not expected for these species to remain in gas phase at the low temperatures of these planets \citep[e.g.,][]{Burrows1999}. The absorption cross-sections are calculated using the methods of \cite{Gandhi2018}.

 The parameter estimation and Bayesian inference is conducted using the Nested Sampling algorithm implemented using PyMultiNest \citep{Buchner2014}. We choose log uniform priors for the volume mixing ratios of all species between -12 and -1. We further expand this prior to $-0.3$ for planets less massive than Saturn ($\sim0.3$ M$_{\text{J}}$) to allow for extremely high ($\sim$50\%) H$_2$O abundances. The temperature prior at the top of the atmosphere is uniform with a lower limit at 0K for $\text{T}_{eq.}<900\text{K}$, $400\text{K}$ for $900\text{K}<\text{T}_{eq.}<1200\text{K}$ and $800\text{K}$ for $\text{T}_{eq.}>1200\text{K}$, and a higher limit at $\text{T}_{eq.}+100\text{K}$. Our retrieval of WASP-19b allows for a wavelength shift relative to the model for VLT FORS2 data only, as performed by \cite{Sedaghati2017}. The retrievals performed in this study have in general 18 free parameters: 7 chemical abundances, 6 for the pressure-temperature profile, 4 for clouds and hazes, and 1 for the reference pressure at the measured observed radius of the planet. 

\begin{deluxetable*}{cc|cccccccc}
\tablecaption{Planetary Properties and Retrieved H$_2$O, Na, and K Abundances (Mixing Ratio) for 19 Exoplanets  \label{table:summary}}
\tablecolumns{6}

\tablewidth{0pt}
\tablehead{
\colhead{$\#$} & \colhead{Planet Name} & \colhead{ M$_{\text{p}}$(M$_{\text{J}}$)} & \colhead{T$_{\text{eq}}$ (K)} & \colhead{log(X$_{\text{H}_2\text{O}}$)} & \colhead{DS$_{\text{H}_2\text{O}}$} & \colhead{log(X$_{\text{Na}}$)} & \colhead{DS$_{\text{Na}}$} & \colhead{ log(X$_{\text{K}}$)} & \colhead{DS$_{\text{K}}$} 
}
\startdata
1   &   K2-18b   & 0.03 & 290    & $-2.36 ^{+ 1.17 }_{- 1.16 }$ & 3.26 & N/A & N/A & N/A &  N/A\\
2   &   GJ3470b   & 0.04 & 693    & $-2.83 ^{+ 0.87 }_{- 0.77 }$  & 3.75 & N/A & N/A & N/A &  N/A\\
3   &   HAT-P-26b & 0.06 & 994   & $-1.83 ^{+ 0.46 }_{- 0.57 }$  & 8.61 & $-9.08 ^{+ 2.04 }_{- 1.88 }$ & N/A & $-10.56 ^{+ 1.21 }_{- 0.96 }$ &  N/A\\
4   &   HAT-P-11b & 0.07 & 831   & $-3.66 ^{+ 0.83 }_{- 0.57 }$  & 4.12 & $-9.36 ^{+ 2.04 }_{- 1.67 }$ & N/A & $-9.80 ^{+ 1.62 }_{- 1.42 }$ &  N/A\\
5   &   WASP-107b & 0.12 & 740    & $-2.87 ^{+ 0.95 }_{- 0.73 }$ & 5.70 & N/A & N/A & N/A & N/A\\
6   &   WASP-127b & 0.18 & 1400   & $-2.13 ^{+ 0.65 }_{- 3.63 }$ & 2.07 & $-2.19 ^{+ 0.81 }_{- 4.41 }$ & 3.66 & $-2.89 ^{+ 0.80 }_{- 3.44 }$ & 4.77\\
7   &   HAT-P-12b & 0.21 & 960    & $-5.70 ^{+ 1.22 }_{- 3.36 }$ & 1.57 & $-8.97 ^{+ 2.33 }_{- 1.94 }$ & N/A & $-8.33 ^{+ 2.32 }_{- 2.09 }$ & N/A \\
8   &   WASP-39b & 0.28 & 1120   & $-0.65 ^{+ 0.14 }_{- 1.83 }$ & 8.92 & $-3.62 ^{+ 1.14 }_{- 2.69 }$ & 3.83 & $-5.62 ^{+ 2.30 }_{- 2.05 }$ & 2.37\\
8   &   WASP-39b$^{*}$ & 0.28 & 1120   & $-2.43 ^{+ 0.27 }_{- 0.24 }$ & 9.20 & $-6.17 ^{+ 0.50 }_{- 0.51 }$ & 3.97 & $-7.24 ^{+ 0.71 }_{- 1.06 }$ & 2.45\\
9   &   WASP-31b  & 0.48 & 1580   & $-4.55 ^{+ 1.77 }_{- 4.33 }$ &  1.65 & $-8.08 ^{+ 2.28 }_{- 2.37 }$ & N/A & $-3.48 ^{+ 1.38 }_{- 2.31 }$ & 2.89\\
10  &   WASP-96b  & 0.48 & 1285   & $-4.95 ^{+ 2.25 }_{- 4.19 }$ & 1.61 & $-5.26 ^{+ 0.75 }_{- 0.59 }$ & 5.16 & $-9.27 ^{+ 1.46 }_{- 1.60 }$ & N/A\\
11   &  WASP-6b   & 0.50 & 1150   & $-8.12 ^{+ 2.52 }_{- 2.41 }$ & N/A & $-8.25 ^{+ 3.10 }_{- 2.30 }$ & N/A & $-3.22 ^{+ 1.21 }_{- 3.79 }$ & 2.55\\
12   &  WASP-17b  & 0.51 & 1740   & $ -3.84 ^{+ 1.27 }_{- 0.51 }$ & 3.36 & $-8.65 ^{+ 1.76 }_{- 1.67 }$ & 1.09 & $-9.62 ^{+ 1.78 }_{- 1.45 }$ & N/A\\
13  &   HAT-P-1b  & 0.53 & 1320   & $-2.54 ^{+ 0.75 }_{- 0.67 }$ & 3.17 & $-8.58 ^{+ 1.20 }_{- 1.79 }$ & 1.32 & $-8.93 ^{+ 2.01 }_{- 1.93 }$ & N/A\\
14  &   HD~209458b & 0.69 & 1450   & $-4.54 ^{+ 0.33 }_{- 0.27 }$ & 6.80 & $-5.47 ^{+ 0.61 }_{- 0.48 }$ & 6.77 & $-7.00 ^{+ 0.59 }_{- 0.49 }$ & 3.90\\
15  &   HD~189733b & 1.14 & 1200   & $-4.66 ^{+ 0.35 }_{- 0.33 }$ & 5.26 & $-4.19 ^{+ 0.67 }_{- 0.73 }$ & 5.51 & $-5.54 ^{+ 0.49 }_{- 0.44 }$ & 3.34\\
16  &   WASP-19b  & 1.14 & 2050   & $-3.43 ^{+ 0.47 }_{- 0.52 }$ & 7.12 & $-5.11 ^{+ 1.00 }_{- 1.05 }$ & 3.48 & $-10.85 ^{+ 1.20 }_{- 0.80 }$ & N/A\\
17  &   WASP-12b  & 1.40 & 2510   & $-3.23 ^{+ 1.42 }_{- 0.80 }$ & 5.41 & $-6.64 ^{+ 2.13 }_{- 2.98 }$ & 1.58 & $-8.94 ^{+ 2.35 }_{- 1.92 }$ & N/A \\
18  &   WASP-43b  & 2.03 & 1440   & $-3.68 ^{+ 0.92 }_{- 0.88 }$ & 3.31 & $-6.95 ^{+ 3.21 }_{- 3.22 }$ & N/A & $-7.50 ^{+ 3.02 }_{- 2.87 }$ & N/A\\
19  &   WASP-33b  & 2.10 & 2700   & $-6.64 ^{+ 3.15 }_{- 3.25 }$ & 1.29 & $-8.98 ^{+ 2.28 }_{- 1.89 }$ & N/A & $-6.77 ^{+ 2.83 }_{- 3.16 }$ & N/A
\enddata
 \tablecomments{The detection significance (DS) is included for each individual species. N/A means that the model without the chemical species had more evidence than the model including it (e.g., $B<1$), that the corresponding species was not included in the model or, in the case of WASP-43b, that it is not possible to provide a DS since no optical data was utilized. Planet mass and equilibrium temperature, with uniform redistribution, are quoted as nominal values; uncertainties in these values are not considered. WASP-39b$^{*}$ corresponds to the case with an upper end of prior on the volume mixing ratios at -1 rather than -0.3.}
\end{deluxetable*}

\begin{figure*}
    \centering
    \includegraphics[width=1.0\textwidth]{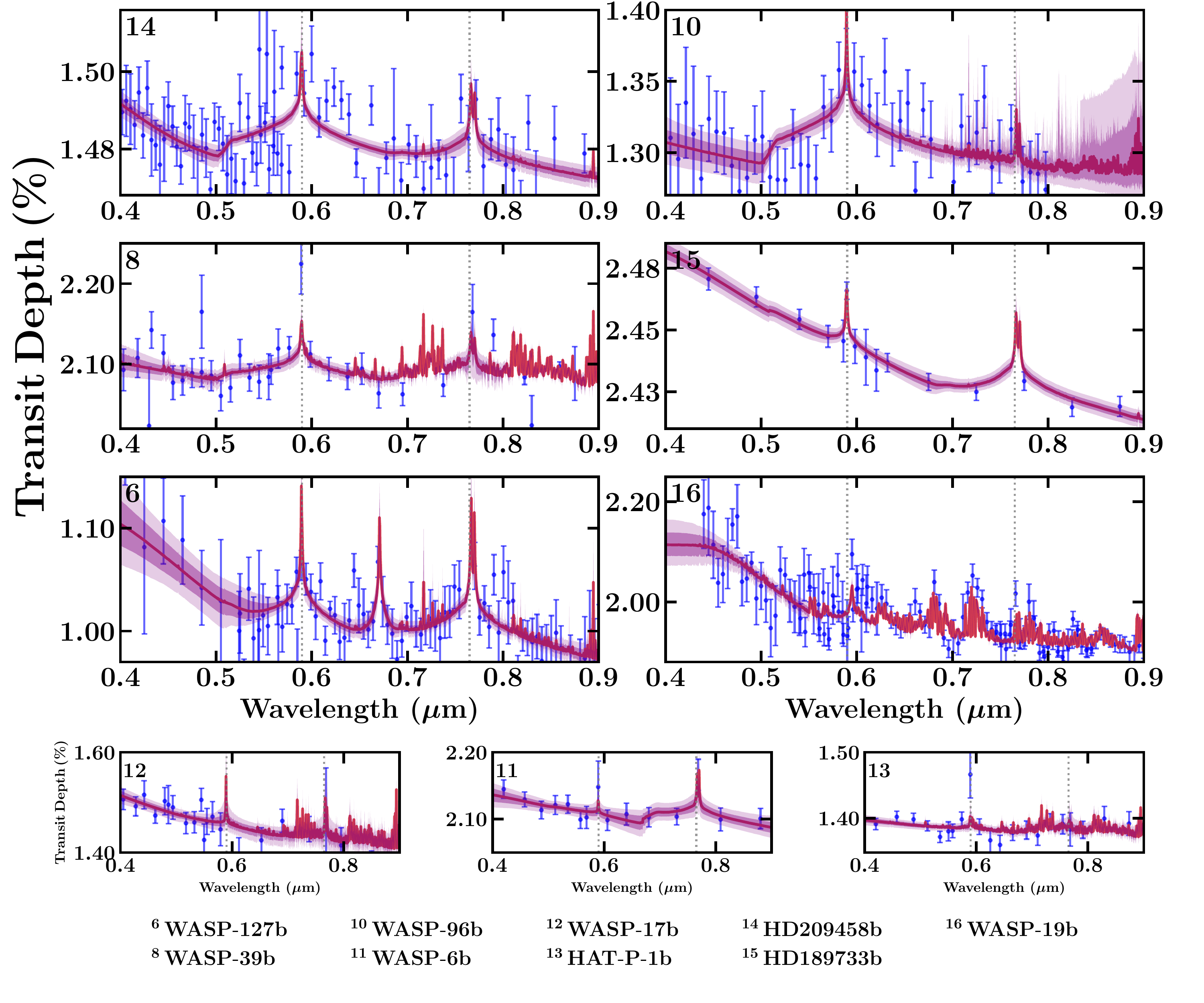}
    \caption{Observations and retrieved model transmission spectra of exoplanets showing evidence of Na and/or K in the optical wavelengths. Dotted lines show the wavelength positions for Na ($\sim0.589\mu$m) and K ($\sim0.770\mu$m). Observations are shown in blue while median retrieved models and confidence intervals (1$\sigma$ and 2$\sigma$) are shown in red and purple respectively. The top six panels show planets with evidence for Na or K above 2$\sigma$, and the bottom three show those with weaker evidence (see Table~\ref{table:summary}). Full optical and infrared spectra are available online at \url{osf.io/nm84s}.}
    \label{fig:composite}
\end{figure*}

\section{Results}
\label{sec:results}

The atmospheric constraints for our sample of 19 transiting exoplanets, ranging from hot Jupiters to cool mini-Neptunes are shown in Table~\ref{table:summary}\footnote{ Priors and posterior distributions are available on the Open Science Framework  \url{osf.io/nm84s}}. The detection significance is calculated from the Bayes factor \citep{Benneke2013, Buchner2014}. We consider reliable abundance estimates to be those with detection significances larger than 2$\sigma$.

\subsection{Abundances of H$_2$O, Na, and K}
\label{subsec:solarestimates}
\begin{figure*}
    \centering
    \includegraphics[width=1.0\textwidth]{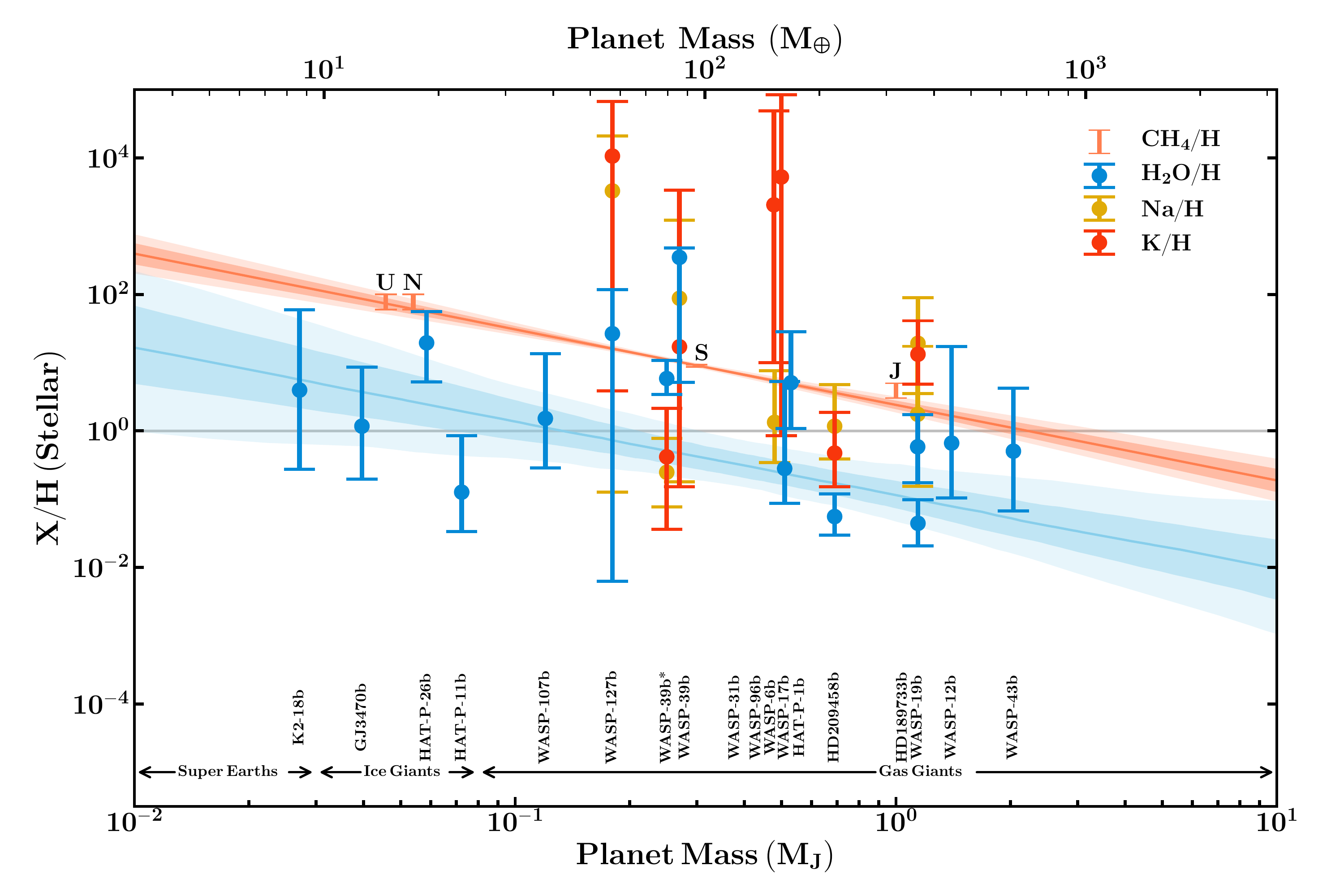}
    \caption{Mass-metallicity relation for planets with chemical detections above 2$\sigma$ significance (see Table~\ref{table:summary}). The H$_2$O/H, Na/H, and K/H abundances are shown in blue, yellow, and orange, respectively. All the abundances are normalized to expectations based on their host stars, as described in Section~\ref{subsec:solarestimates}. The metallicity estimates for the solar system giant planets using their methane (CH$_4$) abundances \citep{Atreya2016} are shown in coral. The coral shaded regions show the 1$\sigma$ and 2$\sigma$ metallicity trend for the solar system planets resulting from a linear fit. The corresponding fit to the exoplanet H$_2$O abundances, excluding WASP-39b (see Section~\ref{subsec:metallicities}), and the confidence intervals are shown in the blue line and sky blue shaded regions. The fits are $\log(\text{CH}_4\text{/H})=-1.11^{+0.11}_{-0.10} \log(\text{M/M}_\text{J})+0.38^{+0.06}_{-0.06} $ and $\log(\text{H}_2\text{O/H})=-1.09^{+0.34}_{-0.33}\log(\text{M/M}_\text{J})-0.95^{+0.21}_{-0.19}$. The alkali (H$_2$O) abundances for the sample are generally above (below) the solar system trend. Results for WASP-39b$^{*}$ (see Table~\ref{table:summary}) and some labels have been offset in mass for clarity.}
    \label{fig:metallicity}
\end{figure*}

We confirm detections of H$_2$O, Na, and K in 14, 6, and 6 planets, respectively, at over 2$\sigma$ confidence, as shown in Table~\ref{table:summary}. Figure~\ref{fig:composite} shows the observed spectra and model fits for planets where at least one of either Na or K is detected. Our retrieved abundances (volume mixing ratios) for these species are broadly consistent with previous surveys and studies on each planet, as discussed in Sections~\ref{sec:intro} and \ref{sec:observations}. The abundances can be assessed relative to expectations from thermochemical equilibrium for solar elemental composition \citep{Asplund2009}. For solar composition (C/O = 0.54), at $\text{T}\gtrsim 1200$ K roughly half the oxygen is expected to be in H$_2$O, at 1 bar pressure \citep{Madhusudhan2012}. Thus, $\log(\text{X}_{\text{H}_2\text{O}})\sim -3.3$ and $-3.0$ for T above and below 1200 K, respectively. Similarly, $\log(\text{X}_{\text{Na}})=-5.76$, and $\log(\text{X}_{\text{K}})=-6.97$ for $\text{T}\gtrsim1100\text{-}1200$ K, below which they enter molecular states \citep{Burrows1999}. Figure~\ref{fig:metallicity} shows the abundances of Na, K, and H$_2$O relative to expectations based on their stellar elemental abundances as described above for solar composition. The stellar abundances \citep[e.g.][]{Brewer2016} used here are available at \url{osf.io/nm84s}. For stars without [Na/H], [O/H], or [K/H] estimates we adopt the [Fe/H] values.

 The best constraints are obtained for H$_2$O across the sample, with precisions between $\sim0.3$ and 1 dex for many of the planets, as shown in figure~\ref{fig:metallicity}. The median H$_2$O abundances for most of the gas giants are substellar, with some being consistent to stellar values within $\sim1\sigma$. On the other hand, smaller planets show an increase in H$_2$O abundances, albeit with generally larger uncertainties. In the ice giants and mini-Neptunes the median abundances are nearly stellar, with the exception of HAT-P-26b and HAT-P-11b, which are significantly superstellar and substellar, respectively. An exception is the hot Saturn WASP-39b for which anomalously high H$_2$O was reported with the latest data \citep{Wakeford2018,Kirk2019}. We find its abundance estimates to be sensitive to the choice of priors; we report two estimates with different priors in Table~\ref{table:summary}. 

Contrary to H$_2$O, the median abundances of Na and K are nearly stellar or superstellar across the seven gas giants. The alkali abundances are retrieved only for the gas giants in our sample, with uncertainties larger than those for H$_2$O.  For planets where they are nearly stellar (e.g., HD~209458b) Na and K are still more enhanced relative to H$_2$O, as with the other planets. The retrieved alkali abundances represent the population of ground state species. While previous studies noted the effect of non-LTE ionization on the absorption strength of alkali lines \citep{Barman2002, Fortney2003}, others \citep[e.g.,][]{Fisher2019} show that the effect is less pertinent for interpretation of low-resolution spectra as in the present study.  Our retrieved Na abundances are consistent with \cite{Fisher2019} for the same planets.  Nonetheless, as discussed in \cite{Welbanks2019}, the simplified model assumptions in \citet[][e.g., isotherms, limited absorbers, and no CIA opacity]{Fisher2019} likely affect the precision and accuracy of their abundance constraints, perhaps explaining their unphysically high temperatures even in LTE. Furthermore, alkali broadening in \cite{Fisher2019} is inconsistent with the more accurate broadening \citep{Allard2019} used in the present work.

\begin{figure*}
    \centering
    \includegraphics[width=1\textwidth]{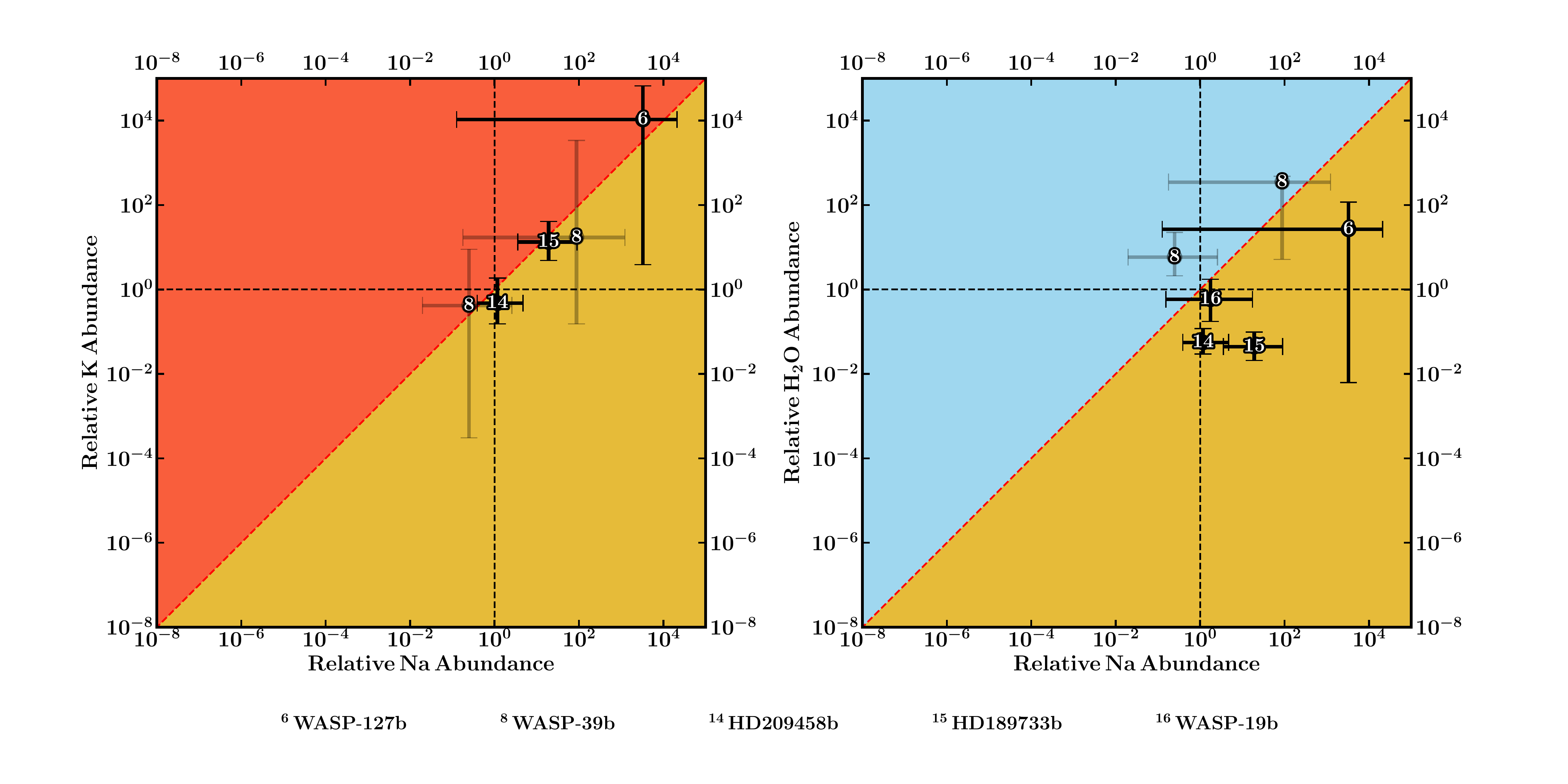}
    \caption{ Normalized abundances of Na, K, and H$_2$O for detections above 2$\sigma$ significance. Left: K vs. Na abundances normalized by host stellar abundances. Right: normalized H$_2$O vs. Na abundances. Abundances are normalized following the description in Section~\ref{subsec:solarestimates}. The red dashed diagonal line in each plot shows the `unity line' where the normalized abundance ratio between species is equal to 1. The black dashed lines show individual normalized mixing ratios equal to 1. Two values are shown for WASP-39b corresponding to Table~\ref{table:summary} in a lighter shade.}
    \label{fig:ratio_plots}
\end{figure*}

\newpage
\subsection{Abundance Ratios and Mass-Metallicity Relation}
\label{subsec:metallicities}

The abundances of Na, K, and H$_2$O provide constraints on their differential enhancements in hot gas giant atmospheres relative to their host stars. Figure~\ref{fig:ratio_plots} shows abundances of different species, relative to their host stars, compared against each other. Planets with significant indications of Na and K, i.e., WASP-127b, WASP-39b, HD~189733b, and HD~209458b, have a normalized K/Na ratio consistent with unity. Thus, the abundances of Na and K closely follow their stellar proportions; they are enhanced or depleted together in the planetary atmosphere relative to the star. On the contrary, the H$_2$O/Na ratios deviate significantly from their stellar expectations, especially for those planets with the tightest constraints: HD~209458b and HD~189733b. The right panel in figure~\ref{fig:ratio_plots} shows that for the small sample of planets with strong detections ($\gtrsim$2$\sigma$) of both Na and H$_2$O, most have preferential enhancement of Na relative to H$_2$O compared to stellar expectations; the only exception being WASP-39b. A similar trend can be inferred for H$_2$O/K, i.e., of K-enhancement and H$_2$O-depletion, given the Na/K ratios consistent with unity (left panel in figure~\ref{fig:ratio_plots}). Given the present small sample of objects (N = 4) with strong detections of all three species (Na, K, and H$_2$O), future observations for more objects are required to further assess the Na/K trends seen here. 

The H$_2$O abundances retrieved across our diverse sample of planets allow us to investigate a mass-metallicity relation for their atmospheres. In the solar system giant planets, CH$_4$ is thought to contain most of the carbon given their low temperatures. Thus, the CH$_4$ abundance has been used as a proxy for the carbon abundance and hence the metallicity \citep{Atreya2016}. A linear fit to the solar system CH$_4$ abundances leads to a `mass-metallicity' relation of $\log(\text{CH}_4\text{/H})=-1.11^{+0.11}_{-0.10}\log(\text{M/M}_\text{J})+0.38^{+0.06}_{-0.06}$. As discussed in Section~\ref{subsec:solarestimates}, the H$_2$O abundances in our exoplanet sample also show a gradually increasing trend with decreasing mass as shown in Fig.~\ref{fig:metallicity}. However, the H$_2$O abundances across the entire sample, down to the mini-Neptunes, largely fall below the solar system metallicity trend based on CH$_4$ abundances. A linear fit to the exoplanetary H$_2$O abundances, excluding WASP-39b due to its strong prior dependence, yields an H$_2$O `mass-metallicity' relation of $\log(\text{H}_2\text{O/H})=-1.09^{+0.34}_{-0.33}\log(\text{M/M}_\text{J})-0.95^{+0.21}_{-0.19}$, which is inconsistent with the solar system CH$_4$ measurements at over 6$\sigma$. On the other hand, the Na and K abundances for the gas giants are mostly consistent with the solar system carbon trend, albeit aided by their larger uncertainties.

\section{Discussion}
\label{sec:conclusions}

Our study reveals three key trends in the atmospheric compositions of our exoplanet sample. Firstly, from mini-Neptunes to hot Jupiters, H$_2$O abundances are generally consistent with or depleted compared to equilibrium expectations based on stellar abundances, and lower than the solar system metallicity trend. Second, the gas giants exhibit Na and K abundances consistent with or higher than those of their host stars and the solar system trend. Lastly, the Na and K elemental ratios are consistent with each other.

The overall low H$_2$O abundances across the sample contrasts with solar system predictions. Besides the carbon enhancements seen in the solar system (Fig.~\ref{fig:metallicity}), other elements such as nitrogen, sulfur, phosphorous, and noble-gases are also enhanced in Jupiter \citep{Atreya2016}; the oxygen abundance is unknown as H$_2$O condenses at the low temperatures of solar system giants. Considering that oxygen is the most cosmically abundant element after H and He, it is expected to be even more enhanced than carbon in giant planets, according to solar system predictions \citep[]{Mousis2012}. Therefore, the consistent depletion of H$_2$O abundances in our sample suggest different formation pathways for these close-in exoplanets compared to the long-period solar system giants. 

The H$_2$O abundance is likely representative of the oxygen abundance (O/H) for our cool Neptunes/mini-Neptunes. The H$_2$O abundance is less sensitive to C/O for T$\lesssim$1200 K, with most of the O bound in H$_2$O regardless of C/O \citep{Madhusudhan2012}. Thus, the H$_2$O abundances for our exo-Neptunes and mini-Neptunes (T$\lesssim$1200 K) indicate somewhat lower metallicities in their atmospheres than solar system expectations \citep{Atreya2016}. On the other hand, for hot gas giant photospheres  (T$\gtrsim$1200 K), the H$_2$O abundance depends on both the O/H and C/O ratios. A C/O $\sim$ 1 can lead to $\sim$100$\times$ depletion in H$_2$O compared to solar C/O (0.54). Thus, even for a solar or super-solar O/H the H$_2$O in hot gas giants can be subsolar if the C/O is high.

The low H$_2$O abundances in hot gas giants are therefore unlikely to be due to low O/H or low overall metallicities, considering the higher alkali enrichments in some planets. An atmosphere depleted up to 100$\times$ in O/H while being enriched in other elements is unlikely \citep{Madhusudhan2014b}. Instead, the H$_2$O underabundance and alkali enrichment are likely due to superstellar C/O, Na/O, and K/O ratios, i.e., an overall stellar or superstellar metallicity but oxygen depleted relative to other species. This addresses the degeneracy between C/O and metallicity prevalent since the first inferences of low H$_2$O abundances in hot Jupiters \citep{Madhusudhan2014a}. 

These H$_2$O abundances in hot gas giants, while generally inconsistent with expectations from solar system and similar predictions for exoplanets \citep[e.g.][]{Mousis2012,Thorngren2016}, could suggest other formation  pathways. The combination of stellar or superstellar metallicities and high C/O ratios in hot Jupiters can instead be caused by primarily accreting high C/O gas outside the H$_2$O and CO$_2$ snow lines \citep{Oberg2011,Madhusudhan2014b}. The high Na and K abundances could potentially be caused by accretion of planetesimals rich in alkalis at a later epoch, potentially even in close-in orbits. Another possibility is the formation of giant planets by accreting metal-rich and high C/O gas caused by pebble drift \citep{Oberg2016, Booth2017}. Future studies could further investigate these avenues to explain the observed trends.

Finally, our study highlights important lessons for the interpretation of atmospheric spectra of gas giants. The differing trends in the abundances of species argue against the use of chemical equilibrium models with the metallicity and C/O ratio as the only free chemical parameters in atmospheric retrievals; different elements can be differently enhanced. Atmospheric models also need to consider the effect of H$_2$ broadening of alkali opacities on abundance estimates. With better quality data and targeted observations, comparative atmospheric characterization of exoplanets as pursued here will likely continue to unveil trends that may inform us about their formation and evolutionary mechanisms. 

\section{Acknowledgements}

LW thanks the Gates Cambridge Trust for support towards his doctoral studies. NM acknowledges support from the Science and Technology Facilities Council (STFC) UK. We thank Dr. Kaisey Mandel for helpful discussions. We thank the anonymous reviewer for their thoughtful comments on the manuscript. This research is made open access thanks to the Bill \& Melinda Gates foundation.

\bibliographystyle{aasjournal}

\appendix

\counterwithin{figure}{section}

\counterwithin{table}{section}

Here we present a subset of the supplementary material included online. Further information, tables, and figures are included in \url{osf.io/nm84s}.

\section{Supplementary information}

\begin{figure}[htbp!]
    \centering
    \includegraphics[width=0.76\textwidth]{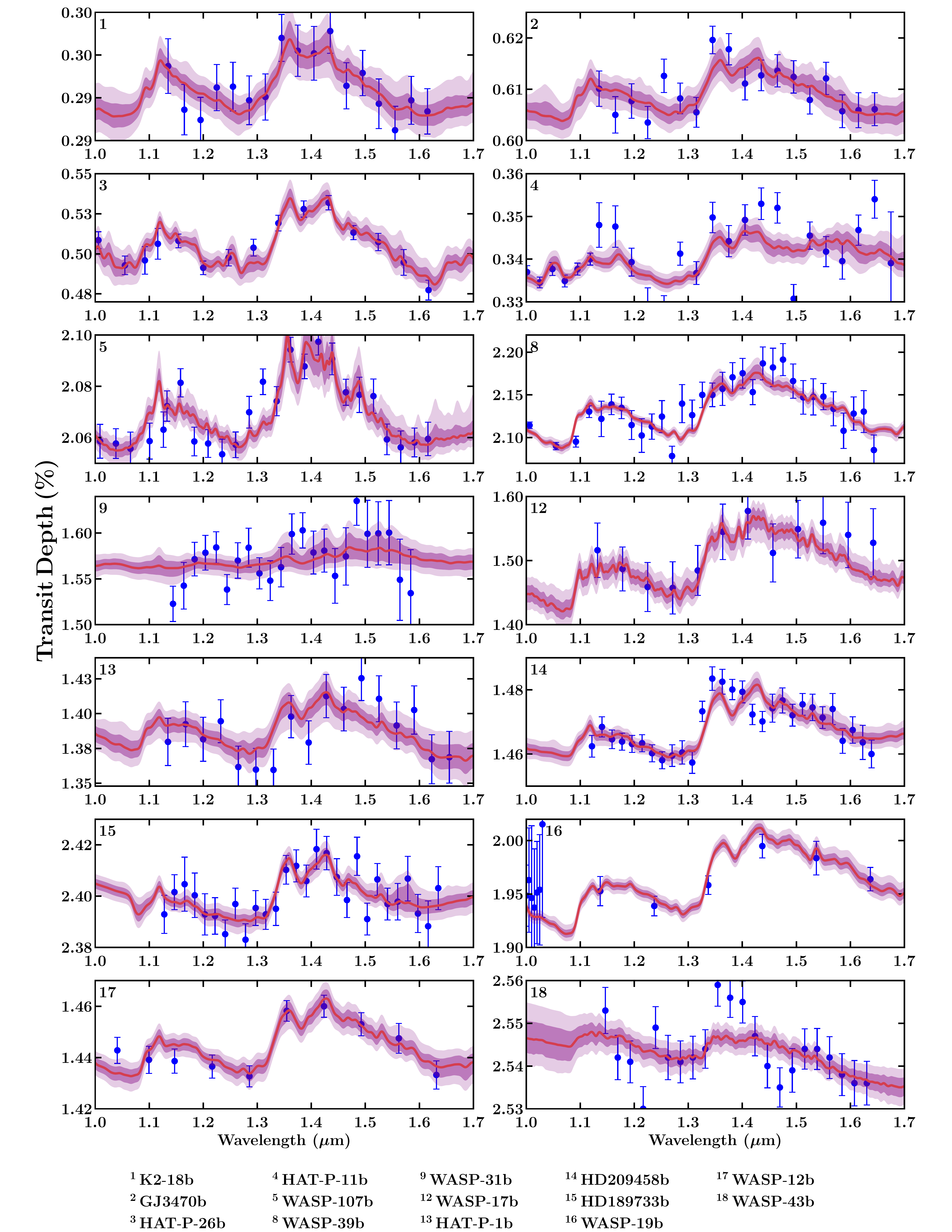}
    \caption{Observations and retrieved model transmission spectra of exoplanets showing evidence of H$_2$O in the near infrared wavelengths. Observations are shown in blue while median retrieved models and confidence intervals (1$\sigma$ and 2$\sigma$) are shown in red and purple respectively. Planets are ordered by increasing mass. All planets shown, except WASP-31b, have a H$_2$O absorption feature with an evidence above 2$\sigma$ (see Table~\ref{table:summary}). These observations were obtained with the HST WFC3 spectrograph. Full optical and infrared spectra are available online at \url{osf.io/nm84s}.}
\end{figure}

\end{document}